\documentclass[9pt,conference]{IEEEtran}

\usepackage[preprint]{waspaa25}

\usepackage{bm} 
\usepackage{lipsum}
\usepackage{multirow}
\usepackage{makecell}

\usepackage{caption}
\usepackage{subcaption}

\title{Task-Specific Audio Coding for Machines:\\
Machine-Learned Latent Features Are Codes for That Machine}

\name{Anastasia Kuznetsova$^{1}$, Inseon Jang$^{2}$, Wootaek Lim$^{2}$, Minje Kim$^{3}$
\thanks{This work was supported by Electronics and Telecommunications Research Institute (ETRI) grant funded by the Korean government
[25ZC1100: The research of the basic media contents technologies].}}
\address{$^{1}$Indiana University, Bloomington, IN, USA\\
$^{3}$Electronics and Telecommunications Research Institute, Daejeon, Korea\\
$^{3}$University of Illinois Urbana-Champaign, Champaign, IL, USA
}

\DeclareMathOperator*{\argmin}{arg\,min}

\begin{document}

\maketitle

\begin{abstract}
Neural audio codecs, leveraging quantization algorithms, have significantly impacted various speech/audio tasks. While high-fidelity reconstruction is paramount for human perception, audio coding for machines (ACoM) prioritizes efficient compression and downstream task performance, disregarding perceptual nuances. This work introduces an efficient ACoM method that can compress and quantize any chosen intermediate feature representation of an already trained speech/audio downstream model. Our approach employs task-specific loss guidance alongside residual vector quantization (RVQ) losses, providing ultra-low bitrates (i.e., less than 200 bps) with a minimal loss of the downstream model performance. The resulting tokenizer is adaptable to various bitrates and model sizes for flexible deployment. Evaluated on automatic speech recognition and audio classification, our method demonstrates its efficacy and potential for broader task and architectural applicability through appropriate regularization. 
\end{abstract}

\section{Introduction}
\label{sec:intro}

Audio codecs have been an active area of research due to their versatile nature. The primary use of  codecs is signal compression with the minimal loss of the perceptual quality for efficient transmission \cite{mp3,PainterT2000ieeeproc,aac-standard, saoc, usac2, amr-wb-standard}. More recently, neural audio codecs \cite{Zeghidour2021soundstream, KumarR2023dac, DefossezA2023encodec} played part in facilitating the success of various generative models, turning inherently continuous audio representations into discrete tokens. Various  speech and audio related tasks have benefited from discrete tokens: speech language models (SLMs), text-to-speech (TTS) \cite{wang2023valle}, voice conversion (VC) \cite{lakhotia2021generative, cui2024recent, Wang2023synth}, automatic speech recogntion (ASR), audio classification (AC) \cite{chen2022beats}, speech enhancement (SE) \cite {Yang2024Interspeech} among others \cite{Mousavi2024dasb, Wu2024CodecSUPERB}. 

Neural audio coding system typically consists of three parts: an encoder that transforms the input audio into a compact (e.g., low-dimensional) feature space, a quantization module that converts the continuous feature vectors into a discrete representation, and the decoder that recovers the original waveform from the de-quantized feature vectors. In other words, if it were not for quantization it is natural to consider it an autoencoder. Indeed, incorporating the quantization module as a trainable part of the model optimization process has been the key to developing a successful neural audio codec. For example, popular neural codecs, e.g., SoundStream~\cite{Zeghidour2021soundstream}, Encodec \cite{DefossezA2023encodec}, and DAC \cite{KumarR2023dac}, are trained using residual vector quantization (RVQ) and guided by reconstruction losses to receive high-fidelity output audio signal. In this scenario, the reconstruction loss is still directly \cite{ZhenK2020spl} or indirectly aiming at retaining the perceptual quality of the original audio, e.g., for speech communication or music streaming. 

However, learning to discretize audio for non-human entities has been underexplored. Discriminative audio tasks, such as ASR or AC, do not require the signal to be reconstructed and perceived by humans, thus fine-grained nuances of audio are redundant and bitrate-inefficient. Instead, codes for a downstream machine learning (ML) task can be further optimized under audio coding for machines (ACoM) paradigm introduced by \cite{mpeg46}. ACoM implies generalized approaches to coding, where the coding efficiency is optimized based on the downstream performance, focusing on the utility of such discrete features for the machines while ignoring its perceptual characteristics. ACoM adheres to a set of principles: a) codes must be efficient in bitrate and size; b) can be used by the machines without degrading the downstream performance; c) need to be optimized for machine consumption \cite{mpeg46}.

One step towards ACoM can be a stream of foundational models, that can be seen as discrete tokenizers via self-supervised learning, such as Wav2Vec2.0 \cite{Baevski2020nips},  WavLM \cite{chen2022wavlm}, or HuBERT \cite{HsuWN2021ieeeacmaslp}. They focus mainly on retrieving the phonetic characteristics of speech input \cite{arora2025slms}. Meanwhile, more general-purpose SSL-based tokenizers, such as BEATs \cite{chen2022beats}, can learn semantic information from general audio as well, improving sound classification tasks. They are designed to learn machine-useful features rather than audio reconstruction, aligning them to the ACoM paradigm. Indeed, the previously mentioned autoencoder-type codecs are also shown to learn codes that are useful for other semantically driven tasks, such as SLMs \cite{wu2024towards, cui2024recent}.



 Likewise those discretization methods tend to focus on accuracy and quality, often overlooking the complexity and bitrate redundancy \cite{Mousavi2024dasb, Wu2024CodecSUPERB, Shi2024ESPnetCodec, guo2025tokenizers}. Moreover, striving for a general-purpose universal tokenizer, the downstream tasks are often trained on frozen tokenizers, leading to performance degradation among various tasks \cite{Mousavi2024dasb} in comparison with the continuous baseline. 

In this paper we introduce an efficient ACoM method that preserves the performance of the downstream model close to non-quantized version of the model. Instead of aiming for learning a universally working embedding space, we propose to repurpose the feature transformation part of any existing neural network-based downstream model as a tokenizer, while the remainder still performs the supervised task on the token input. The split is useful when part of the downstream task can be divided into two networked entities, e.g., a user device and the cloud server, where the trade-off between the computational cost and representation quality starts to matter. The proposed method finetunes an existing neural network with a guide by a task-specific loss function along with the RVQ losses, ensuring a task-specific bitstream as an intermediate, transmission-friendly feature representation. The proposed method can discretize any internal feature layer output, providing flexibility to beat the desired quality versus model complexity trade-off. 

Supervised learning of discrete tokens is not new: in \cite{du2024cosyvoiceAS}, an ASR model is used to train a discrete tokenizer, which turns out to be useful for a subsequent TTS system. However, it does not focus on the tokenizer's complexity and compression efficiency. On the other hand, BEATs \cite{chen2022beats} for audio classification presents low bitrate, but applies a multi-stage iterative approach to tokenizer pretraining, which is not always convenient to optimize in practice. In this paper, we show that RVQ-based quantization with proper regularization can serve as means for efficient compression. We evaluate the proposed approach using two popular downstream tasks: ASR and AC, but our method is expandable to a variety of tasks and is architecture independent provided it is optimized with the appropriate regularization. 


\section{Method}
\label{sec:method}

Conventional ML pipelines that involve both on-device signal acquisition and cloud computing consist of several stages as shown in Figure \ref{fig:conventional_pipeline}. First, the audio received from the microphone on device is compressed using a conventional DSP-based or a neural codec. Following compression, the signal is sent over the network to the cloud, where it is decoded and further processed by the downstream ML model for prediction of the target variables (e.g., class labels). The conventional approach contains several redundancies that can be eliminated: (a) conventional codecs are designed for speech/audio communication considering perceptual quality, and are unaware of the downstream task, resulting in unnecessarily higher bitrates or sacrificed performance on the downstream task (b) neural codecs tend to be too heavy to be deployed in the device, especially if the quality of the code matters (c) given that the downstream model needs to do feature transformation again, the encoding and decoding processes are computationally redundant. 

As an alternative, we propose a task-specific, machine oriented codec, which eliminates the necessity of a standalone audio codec from the ML pipeline. Instead, we propose to repurpose the earlier part of the ML model to transform the raw input signal into a compact feature space, which is what those layers are doing anyway, and then introduce quantization in that learned feature space. 

The proposed approach offers the following benefits. First, we reduce the bitrate of the system's encoder, increasing the transmission speed. The quantization process is guided with the task-specific loss, which specializes discrete tokens discarding irrelevant information (e.g., speaker characteristics from the ASR pipeline) as well as eliminates the need for using extra bits of information. Second, the ML pipeline can save the cost of running a codec entirely, as the encoder is replaced by the existing downstream ML model's layers. Decoding is unnecessary in this pipeline, too.  
Finally, the proposed scheme is universal and transferrable to many speech and audio tasks. It can accommodate a variety of resource constraints on the device side, allowing for different space constraints while preserving reasonable trade-off between the downstream performance and space. 

\subsection{Proposed Pipeline}

\begin{figure}
     \centering
     \begin{subfigure}[b]{0.4865\columnwidth}
         \centering
         \includegraphics[width=\textwidth]{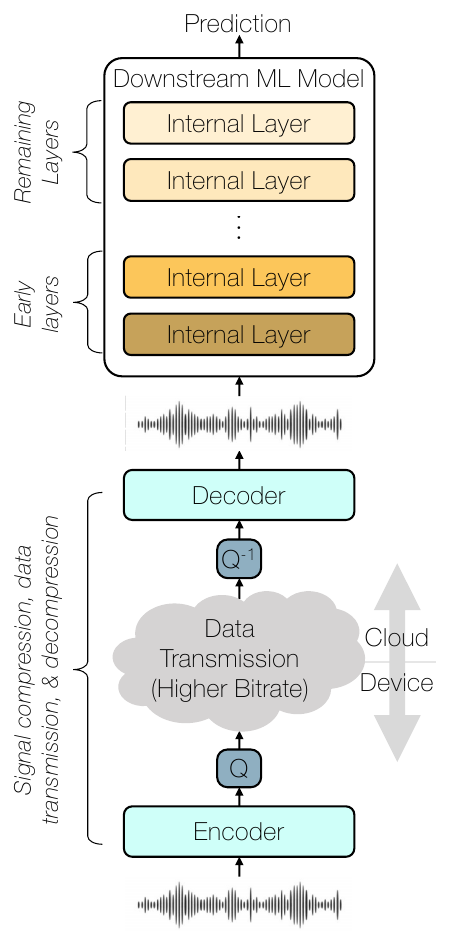}
         \caption{}
         \label{fig:conventional_pipeline}
     \end{subfigure}
     \hfill
     \begin{subfigure}[b]{0.4935\columnwidth}
         \centering
         \includegraphics[width=\textwidth]{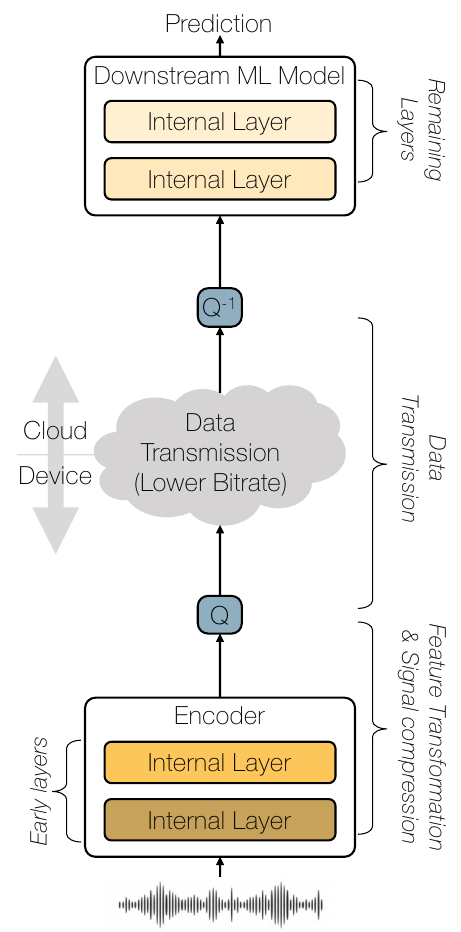}
         \caption{}
         \label{fig:proposed_pipeline}
     \end{subfigure}
        \caption{(a) The conventional ML pipeline involving a codec. The ML model in the cloud is tasked to decompress the signal, perform feature transformation, and the downstream task. (b) The proposed feature coding approach that offloads a few early layers of the downstream model to the device for feature transformation. Trainable quantization on the task-specific feature space ensures more coding gain at a similar downstream performance.}
\end{figure}

Figure \ref{fig:proposed_pipeline} shows the outline of the proposed approach. Here, the main assumption is that the cloud ML model can be split into two parts, the earlier module and remaining part. For example, in a neural network-based model, the first few layers can be offloaded to the edge device for processing, while the remainder reside in the cloud. This kind of split model can be preferred in various use cases, such as when the private user data is preferred to be processed on the device, in order to reduce cloud computing cost, etc. At any rate, by being able to split the model, the entire ML pipeline can flexibly adapt to a particular test-time user environment. 

Based on this assumption, we propose to remove the codec part in the conventional scenario. Instead, the proposed ACoM scenario can push some of the early layers of the ML model to the device for processing, which essentially works like the encoder of a codec. Then, the input audio is transformed into a feature vector by $M$ layers of the encoder. Residual vector quantization (RVQ) \cite{Zeghidour2021soundstream} follows to quantize the feature vector into trained codewords, whose codeword assignment index is sent over to the cloud as a discrete token, Subsequently, the remaining layers in the cloud are optimized to retrieve information from the dequantized version of the feature vectors. Finally, the downstream ML model performs the original prediction task bearing with the potential quantization error introduced by the RVQ module. 

Formally, let $\bm x \in \mathbb{R}^{T}$  be raw audio input to the encoder, where $T$ is the number of audio samples. In total the ML model $\mathcal{F}(\cdot)$ contains $L$ layers. Let $\mathcal{F}_i(\cdot)$ represent the $i\text{-th}$ layer, for $i = 1, 2, \dots, M$ hosted on device, while $L-M$ layers are hosted in the cloud. Let $\bm h_i = \mathcal{F}_i(\bm x)$ denote the output of the $i\text{-th}$ layer. Then, the on-device encoder is defined as a cascade of $M$ layers: $\mathcal{F}_M\circ\mathcal{F}_{M-1}\circ\cdots\circ\mathcal{F}_1(\bm x)$. 



\subsection{Quantization Mechanism}
Subsequently, encoder's output $\bm h_M$ is quantized using RVQ \cite{Zeghidour2021soundstream}. To this end, we define a series of $K$ codebooks, each of which has $V$ $D$-dimensional codewords: $\bm C\in\mathbb{R}^{D\times V\times K}$. 
The first stage of the RVQ process replaces the encoder output $\bm h_M$ with the closest code word found in the first codebook $\bm C_{:,:,1}$ as follows:
\begin{equation}
    \bm h_M \approx \bm C_{:,v^*_1,1}, \text{ where } v^*_1=\argmin_v\|\bm h_M - \bm C_{:,v,1}\|_2.
\end{equation}
where $v^*_k$ is the index of the best matching codeword within the $k$-th codebook. Then, the first residual $\bm r_1$ in the code space is calculated, $\bm r_1=\bm h_M-\bm C_{:,v^*_1,1}$, to define the new input to the second RVQ stage:
\begin{equation}
    \bm r_1 \approx \bm C_{:,v^*_2,2}, \text{ where } v^*_2=\argmin_v\|\bm r_1 - \bm C_{:,v,2}\|_2.
\end{equation}
There are $K$ such stages in RVQ to collectively reconstruct the original feature vector $\bm h_M\approx\hat{\bm h}_M=\sum_{k=1}^K \bm C_{:,v^*_k,k}$. We denote this RVQ process by $\mathcal{Q}(\bm x)$, which results in $K$ indices to the closest codewords for $K$ codebooks, respectively: $\bm v^*=[v^*_1, v^*_2, \ldots, v^*_K]=\mathcal{Q}(\bm x)$. In this way, instead of sending $\bm h_M$ to the remaining layers in the cloud, the transmission of the discrete token $\bm v^*$, suffices for the approximation. The rest of the cloud layers proces the dequantized features as input: $\hat{\bm y}=\mathcal{F}_L\circ\mathcal{F}_{L-1}\circ\cdots\circ\mathcal{F}_{M+1}(\hat{\bm h}_M)$, where $\hat{\bm y}$ is the prediction of the ground-truth target ${\bm y}$. 



The empirical bitrate of the proposed model depends on the number of codebookk $K$, the size of each codebook $V$, and finally, the distribution of codeword assignment. Let $R$ be the frame rate, then the raw bitrate (i.e., before considering the entropy of the code distribution) can be calculated as follows: $\text{Bitrate}_\text{raw} = R \cdot K \cdot \lceil \log_2 V\rceil$.



Finally, coding efficiency of the proposed approach can be further improved via entropy coding, e.g., Huffman coding as shown in the soft-to-hard quantization method \cite{AgustssonE2017softmax}, where the entropy of the codes was regularized by a loss term. In this work, instead of controlling the entropy directly, we adopt the \textit{codebook utilization} concept proposed in DAC \cite{KumarR2023dac}, where codebook under-utilization corresponds to low empirical entropy. In DAC, there was no specific use of entropy coding, but we bring this concept back to our model, associating it with codebook utilization during our experiments. Eventually, our method benefits from entropy coding, e.g., when a codebook is under-utilized, we assume that entropy coding can follow up for a further bitrate reduction, leveraging the low-entropy code distribution. Hence, we report entropy-based bitrates as the model's compression performance. To this end, we first calculate the frequency of each codeword of the $k$-th codebook given the entire test samples $\{\bm x^{(i)}\}_{i=1}^N$:
\begin{equation}
    \rho_v=\frac{1}{N}\sum_{i=1}^N \mathcal{I}({v^*_k}^{(i)}=v), \quad \forall v=\{1,\ldots,V\},
\end{equation}
where $\mathcal{I}(\cdot)$ is the indicator function. Then, the frequency vector $\bm \rho$ is used to calculate the empirical entropy of the $k$-th codebook $\mathcal{H}(\bm C_{:,:,k})$.
\begin{equation}
    \mathcal{H}(\bm C_{:,:,k})=-\sum_{v=1}^V \rho_v\log \rho_v.
\end{equation}
Eventually, the final entropy of all codebooks is the sum of their empirical entropy values: $\mathcal{H}(\bm C)=\sum_{k=1}^K\mathcal{H}(\bm C_{:,:,k})$. The empirical entropy is the lower bound of a Huffman coder's bitrate for a frame. Considering the frame rate, the final bitrate is bounded by 
\begin{equation}\label{eq:h_bitrate}
    \text{Bitrate}_\text{ent}\geq R\cdot \mathcal{H}(\bm C).
\end{equation}



\section{Experimental Setup}
\label{sec:experiment}

We empirically show the coding efficiency and model complexity reduction properties of the task-specific quantization method. First, we lower the bitrate of the proposed models as much as possible while preserving the quality of the continuous baseline model. Second, we assess the trade-off between the coding efficiency, the downstream performance, and computational complexity. In particular, we define our ``encoder" flexibly by varying the number of the on-device layer $M$ to investigate this trade-off: a small $M$ makes on-device processing more affordable at the cost of suboptimal coding gain. On the contrary, a large $M$ lets the quantizer work in a more abstract (i.e., easier to compress) feature space, although it could increase the on-device encoder complexity accordingly. 
An alternative training strategy is to use discrete RVQ tokens as input features to encoder-decoder model. However, in that case we would have to retrain the whole system from scratch which is computationally inefficient compared to the quantization of fully trained model, and thus defeats the purpose of the proposed method.
Additionally, we use DAC \cite{Kumar2023arxiv} to simulate Fig. \ref{fig:conventional_pipeline}, whose reconstruction is fed to the continuous baseline models to show that the conventional pipeline may not be optimal both in terms of computational efficiency and quality.

We establish the universality of the proposed model quantization approach by experimenting with speech recognition (ASR) and audio classification (AC) systems, showcasing its effectiveness on both speech data (ASR) and general audio discriminative tasks (AC).

\subsection{Models}

For ASR, we chose a Conformer-Transformer architecture with 12 encoder and 6 decoder layers as defined in SpeechBrain toolkit \cite{speechbrain_v1}. The continuous model serves as a baseline, which is also used to load the pretrained weights that are later finetuned along with the RVQ loss functions. The ASR model is trained using 5,000 BPE subwords \cite{Kudo2018Sentencepiece}, Connectionist Temporal Classification (CTC) \cite{graves2006icml} loss as well as KL-divergence as a label smoothing loss, guided by Adam optimizer \cite{Kingma:2015}. In the quantized version of the ASR model, we add codebook and commitment losses as regularizers as in the original VQ-VAE method \cite{OordA2017vqvae} that the SoundStream model \cite{Zeghidour2021soundstream} used, thus the total ASR loss is defined as (\ref{eq:asr_loss}), where $\lambda$ and $\beta$ are the coefficients for ASR-related losses and the RVQ regularizer, respectively.
\begin{equation}
\label{eq:asr_loss}
    \mathcal{L}_\text{ASR} = \lambda\mathcal{L}_\text{CTC} + (1 - \lambda)\mathcal{L}_\text{KL} + \beta(\mathcal{L}_\text{code} + \mathcal{L}_\text{commit})
\end{equation}

As the baseline for the AC task, we also use SpeechBrain's ECAPA TDNN \cite{desplanques2020ecapa} architecture with dilated convolutions, squeeze and exitation blocks, and attentive statistical pooling. The main objective is the additive angular margin softmax loss \cite{xiang2019margin} accompanied by the codebook loss regularizer as in Eq. \ref{eq:asr_loss}. The model contains 4 encoder layers followed by attentive statistical pooling and a linear classifier. Before applying RVQ, we apply average pooling across the time dimension to lower the original frame rate $R=160$ to $40$.

We perform hyperparameter optimization using the tree-structured Parzen estimator (TPE) algorithm \cite{watanabe2023tree, xavier_bouthillier_2022_0_2_6}. For ASR, we tune batch sizes, learning rates, and the code vector dimension $D$. For AC, we additionally tune the $\beta$ hyperparameter for quantizer regularization.

\subsection{Data} For ASR experiments, we use the conventional LibriSpeech \cite{PanayotovV2015Librispeech} dataset containing 960h of training data with a maximum audio length of 10 seconds. Only \textit{train-clean-100} is used for hyperparameter search.

For AC UrbanSound8k dataset \cite{Salamon2014urbansound} is used. It contains 8,732 sound excerpts $\leq4$ seconds long that are annotated for 10 sound classes, such as air conditioner, gun shot, siren, children playing, etc. We use the original 10-fold split and perform cross-validation as suggested by authors. Both datasets are sampled at 16 kHz.

\begin{table}[t]
\caption{Resuts for ASR on LibriSpeech \cite{PanayotovV2015Librispeech}  with different depths of encoder quantization. GMACs $\downarrow$ are reported to measure the on-device portion of the processing pipeline. in comparison to conventional coding approach based on DAC quantization. We only compute MACs for the DAC bitrates corresponding to quantized ASR models' bitrates.
WER $\downarrow$ is reported on \textit{test-clean} and \textit{test-other} subsets.}
\setlength{\tabcolsep}{0.3em}
\centering
\resizebox{\columnwidth}{!}{%
\begin{tabular}{c|c|c|c|c|c|c|c|c}
\toprule
\makecell{\textbf{No.}\\\textbf{Codebooks}} & \makecell{\textbf{Quant.} \\ \textbf{layer}} & \makecell{\textbf{Codebook} \\ \textbf{size\&dim}} & \makecell{$\textbf{BR}_{\text{raw}}$ \\ \textbf{(bps)}} & \makecell{\textbf{WER}\\\textbf{(test-}\\ \textbf{clean)}} & \makecell{\textbf{WER}\\\textbf{(test-} \\ \textbf{other)}} & \makecell{\textbf{Entropy}\\\textbf{(frame)}} & \makecell{$\textbf{BR}_{\text{ent}}$ \\ \textbf{(bps)}} & \makecell{\textbf{GMACs}\\ \textbf{(on-} \\ \textbf{device)}} \\
 \midrule
\makecell{Cont. \\ baseline} & - & -  & - & 2.01 & 4.52 & - &  -& - \\
\midrule 
\makecell{Cont.\\ DAC 250} & - &  - & - & 2.94  &19.91& 44.99 & 174.68 & 12.30\\
\midrule
\makecell{Cont.\\DAC 500} & -& - & - & 2.94  & 4.53 & 13.99 & 730.74 & 12.30 \\
\midrule
12 & 4 & 1024, 512 & 3000  & 4.06 & 10.1 & 4.76 & 1428.30  & 2.85 \\
12 & 6 & 1024, 512 & 3000  & 3.04 & 7.28 & 5.14 & 1542.45  & 4.22 \\
12 & 8 & 1024, 512 & 3000  & 2.87 & 6.79 & 5.21 & 1563.36  & 5.48 \\
\midrule
2 & 4 & 1024, 64 & 500  & 2.25 & 5.33 & 9.98 & 499.06 & 2.95 \\
2 & 6 & 1024, 64 & 500  & 2.23 & 5.01 & 10.85 & 542.52 & 4.32 \\
2 & 8 & 1024, 64 & 500  & \textbf{2.21} & \textbf{4.99} & 10.95 & 547.59 & 5.69 \\
\midrule
1 & 4 & 8192, 8 & 325  & 2.87 & 7.53 & 6.38 & 159.41  & 2.75 \\
1 & 6 & 8192, 8 & 325  & 3.21 & 8.23 & 5.76 & 144.12  & 4.12 \\
1 & 8 & 8192, 8 & 325  & 3.03 & 7.33 & 6.05 & 151.33   & 5.49 \\
\midrule
1 & 4 & 1024, 64 & 250 & 2.93 & 7.66 & 5.34 & 133.49 & 2.75  \\
1 & 6 & 1024, 64 & 250  & 2.9 & 7.34 & 5.27 & 131.68 & 4.12 \\
1 & 8 & 1024, 64 & 250  & 2.79 & 6.89 & 5.26 & \textbf{131.62}  &  5.49\\
\bottomrule
\end{tabular}}
\label{tab:result_asr}
\end{table}

\subsection{Evaluation}

The quality of the ASR system is measured in the word error rate (WER) reported on \textit{test-clean} and \textit{test-other} subsets of LibriSpeech \cite{PanayotovV2015Librispeech}. Note that the lower WER is the better, which we denote by $\downarrow$ in the result tables. The performance of the AC is assessed via classification accuracy (Acc), which is better if it is higher, i.e., $\uparrow$. To evaluate the coding gain after entropy coding we compute entropy based bitrate ($\text{Bitrate}_\text{ent}, \downarrow$) defined in Eq. (\ref{eq:h_bitrate}).
Additionally, to analyze model complexity, we compute the number of giga multiply-accumulate operations (GMAC) for a one-second audio sample using PyFlops\footnote{\url{https://github.com/sovrasov/flops-counter.pytorch/tree/master}}. We compute GMACs $\downarrow$ of our $M$ encoder layers as well as the RVQ modules to compare them with the 16 kHz version of DAC's encoding, RVQ, and decoding processes, as a simulation of the conventional scenario when the audio needs to go through the neural codec first (Fig. \ref{fig:conventional_pipeline}). We also report raw and entropy-based bitrates for both ASR and AC datasets.

\section{Results}
\label{sec:results}


Table \ref{tab:result_asr} presents WERs for the ASR task on the \textit{test-clean} and \textit{test-other} subsets of LibriSpeech, along with the corresponding bitrates, the on-device portion of the model's GMACs, and corresponding quantization configurations. The table compares a continuous baseline ASR model with models with different quantization strategies. As a simulation of Fig. \ref{fig:conventional_pipeline} scenario, we finetune an ASR model using waveforms reconstructed by a DAC at 250 bps and 500 bps, to evaluate the neural codec's effects on the ASR performance. In this case, DAC's encoder and its RVQ routine is the only on-device operations, amounting to a static 12.3 GMACs.


The continuous baseline ASR model achieves WERs of 2.01 and 4.52 on \textit{test-clean} and \textit{test-other} respectively, which is the performance upper bound of any systems involving quantization. Retraining the same continuous model on DAC reconstructed waveforms at 250 bps leads to a degradation in performance: 2.94 for \textit{test-clean} and 7.58 for \textit{test-other} at 250 bps. At a higher rate of 500 bps, the WERs are similar (2.94 and 7.12), which deems the perceptual quality improvement inefficient for ACoM. 

The proposed method (Fig. \ref{fig:proposed_pipeline}) introduces RVQ to any selected ASR encoder layer instead of a separate waveform coding. In doing so, we also tested various hyperparameters that affect the compression ratio and the loss of information. As for the number of codebooks that linearly increases the raw bitrate, we found that two codebooks are enough for the ASR task, while 12 codebooks are adding additional burden to the optimization process, resulting in lower ASR performance. Of the three chosen encoder layers, 4th, 6th, and 8th, we found a steady trend that the quantization in the higher layer results in better performance (4.99 in 8th), especially for the noisy test set \textit{test-other}. This improvement comes at the cost of increased on-device run-time complexity from 2.95 to 5.69 GMACs. Overall, we observe the best WERs, 2.21 and 4.99, at the 8th layer with two codebooks, that are close to the continuous baseline's. Compared to the Fig. \ref{fig:conventional_pipeline} pipeline, ours achieves much better WERs, especially on the noisy input (7.12 vs. 4.99), even without the computational overhead (12.3 GMACs) that the DAC module comes with. 

The raw bitrate of the two-codebook quantizers is 500 bps, which is already significantly low. In theory, we can further compress the bitstream via an entropy coding scheme, although in this particular case, due to the relatively high per-frame entropy, entropy coding did not result in additional compression. However, it is worth mentioning the very low entropy bitrates of one codebook cases, that achieve 131.62 bps, reaching the theoretical bound of speech communication, 100 bps \cite{vankyuk2017information}. The ultra-low bitrate coding scheme sacrifices the WER performance, accordingly (2.79 and 6.89), although they could be an acceptable performance considering the amount of bitrate saving. In addition, they are still a better solution than the DAC and the continuous ASR pipeline in terms of both WER and complexity.

\begin{table}[t]
\caption{Results for cudio classification (Acc $\uparrow$) on UrbanSound8k \cite{Salamon2014urbansound}. The accyracy Acc $\uparrow$ is reported on the provided test set. The second row Cont. DAC 750 shows the result for DAC reconstructed signals at 750 bps, closest to 800 bps quantized AC system.  
}
\setlength{\tabcolsep}{0.3em}
\centering
\resizebox{\columnwidth}{!}{%
\begin{tabular}{c|c|c|c|c|c|c|c}
\toprule
\makecell{\textbf{No.}\\\textbf{Codebooks}} & \makecell{\textbf{Quantized} \\ \textbf{layer}} & \makecell{\textbf{Codebook} \\ \textbf{size\&dim}} & \makecell{$\textbf{BR}_{\text{raw}}$ \\ \textbf{(bps)}} & \makecell{\textbf{Test}\\ \textbf{Acc}} & \makecell{\textbf{Entropy}\\\textbf{(frame)}} & \makecell{$\textbf{BR}_{\text{ent}}$ \\ \textbf{(bps)}} & \makecell{\textbf{MMACs}\\ \textbf{(on-device)}} \\
\midrule
\makecell{Cont.\\baseline} & - & - & - & 0.797 & - & - & -  \\
\midrule
\makecell{Cont.\\DAC 250} & - & - & - & 0.591 & 6.23 & 155.86 & 12,300\\
\midrule
\makecell{Cont.\\DAC 750} & - & - & - & 0.749 & 20.37 & 1528.95 & 12,300 \\
\midrule
2 & 1 & 1024, 8 & 800 & 0.761 & 11.84 & 947.02 & 42.34 \\
2 & 2 & 1024, 8 & 800 & 0.762 & 11.55 & 924.00 & 287.02 \\
2 & 4 & 1024, 64 & 800 & 0.782 & 3.14 & 251.44 & 784.99\\
\midrule
1 & 1 & 1024, 8 & 400 & 0.761 & 5.62 & 224.60 & 41.7 \\
1 & 2 & 1024, 8 & 400 & 0.758 & 5.43 & 217.25 & 286.38 \\
1 & 4 & 1024, 512 & 400 & 0.793 & 9.92 & 396.56 & 814.46\\
\midrule
1 & 1 & 32, 8 & 200 & 0.694 & 3.50 & 139.93 & 41.5\\
1 & 2 & 32, 8 & 200 & 0.743 & 2.71 & 108.53 & 286.18 \\
1 & 4 & 32, 512 & 200 & 0.802 & 4.21 & 168.44 & 801.76\\
\bottomrule
\end{tabular}
}
\label{tab:result_ac}
\end{table}


Audio classification results outlined in Table \ref{tab:result_ac} confirm our observations. Our quantizer with one codebook, when applied to the 4th classifier layer, results in slightly higher than the continuous baseline classification accuracy (80.2 vs. 79.9\%). Given the raw bitrate of 200 bps, after entropy coding the lower bound goes down to 168.44 bps. As opposed to ASR results, continous baseline trained from the DAC reconstructed audio at 750 bps does not deteriorate the quality of the AC much. However, there still exists a significant gap in computational efficiency due to DAC's computational overhead. 

Overall, we observe the tradeoff between the choice of quantization layer, bitrates, and classification accuracies, which is a design choice depending on how much computation the device can afford. However, the overall trend is that the classification performance can be maintained, while the coded feature representations can be extremely compressed, which is a huge advantage in split architecture.

\section{Conclusion}
\label{sec:conclusion}

We explored task-specific quantization strategy that improved computational efficiency of encoder-decoder tasks (ASR and AC) under the ACoM paradigm. Leaving out the factor of human perception in the training allowed to balance quality and computational complexity of the models via making the RVQ based representations more suitable for machine `understanding'. Splitting the model between the device and the cloud offers flexibility for edge devices of various constraints, and showed the potential for achieving ultra-low bitrates while preserving reasonable performance. Source codes can be found at: \url{https://minjekim.com/research-projects/acom#waspaa2025}.

\bibliographystyle{IEEEtran}
\bibliography{anakuzne, mjkim}

\end{document}